\documentclass[utf8]{FrontiersinHarvard} 

\usepackage{url,hyperref,lineno,microtype,subcaption}
\usepackage[onehalfspacing]{setspace}
\usepackage[T1]{fontenc}
\usepackage{caption}
\usepackage{subfig}
\usepackage{booktabs}

\providecommand{\helveticabold}{\bfseries\sffamily}
\def\keyFont{\fontsize{8}{11}\helveticabold}
\def\firstAuthorLast{Savvidou, S.} 
\def\Authors{Sofia Savvidou\,$^{1,2}$}
\def\Address{$^{1}$Space Telescope Science Institute, 3700 San Martin Drive, Baltimore, MD 21218, USA

$^{2}$Max-Planck-Institut f\"ur Astronomie, K\"onigstuhl 17, 69117 Heidelberg, Germany}
\def\corrAuthor{Sofia Savvidou}

\def\corrEmail{ssavvidou@stsci.edu}

\begin{document}
\onecolumn
\firstpage{1}

\title[A giant solution to the disk mass budget problem of planet formation]{A giant solution to the disk mass budget problem of planet formation} 

\author[\firstAuthorLast]{\Authors} 
\address{} 
\correspondance{} 

\extraAuth{}

\maketitle

\begin{abstract}

\section{Understanding how dust evolves in protoplanetary disks is crucial to constraining the initial conditions of planet formation. The apparent "mass budget problem", which stems from the comparison of the observed disk masses to the ones inferred for exoplanets, remains debated, as it is unclear whether the discrepancy arises from limitations in interpreting disk observations, from evolutionary processes that rapidly deplete dust, or from incorrect assumptions about the initial disk mass distribution. This work is build on the analysis presented in \citet{SavvidouBitsch2025} by separating the cumulative distribution functions of dust masses at different evolutionary stages into different populations according to the initial disk masses and embryo injection times. The best match to observations comes from disks with intermediate initial disk masses around 4-7\% $M_{\odot}$. The largest discrepancy between the total dust mass in the models and the estimated through an "optically thin" approximation comes from the models that have the most favorable conditions for giant planet formation and thus contain a large fraction of giants and subsequently trapped "optically thick" dust mass because of the pressure bumps they generate. However, the final dust masses remain higher compared to the estimates from the observed evolved disks. Example cases in this work including planetesimal formation show that the pressure bumps that giant planets form can be prime locations for planetesimal formation and the conversion to planetesimals significantly decreases the dust mass, as expected. However, (giant) planet formation is not influenced showing that the mass in evolved protoplanetary disks can be estimated to be quite low but it can be a natural consequence of planetesimal and planet formation along with depletion due to radial drift.}

\tiny
 \keyFont{ \section{Keywords:} protoplanetary disks -- circumstellar matter -- dust evolution -- planet formation -- methods: numerical} 
\end{abstract}

\section{Introduction}

The mass budget problem was introduced and first solutions were discussed in the seminal papers of \citet{GreavesRice2010,NajitaKenyon2014,Mulders+2015,Mulders+2018} and \citet{Manara+2018} after comparing the estimated masses of protoplanetary disks observed in the millimeter to those of known exoplanets, especially giant planets at wide orbits, and finding that the latter tend to be at best equal to or often higher than the former. It was already discussed in these early works that this discrepancy could be attributed to some degree to the underestimation of the disk masses due to uncertainties in the opacities and temperatures in the flux-to-mass conversion \citep{Hildebrand1983}. Beyond this, it was discussed that more massive disks could be found at earlier evolutionary stages, hence Class II disk could be the remnants of planet formation rather than the precursors. Indeed, large surveys of Class 0/I systems indicate not only that their dust masses are higher than Class II systems in the same star-forming region (SFR) but also that masses estimated from VLA observations are significantly higher than masses estimated from dust continuum observations with the Atacama Large Millimeter Array (ALMA) \citep{Tychoniec+2020}. However, there are still questions in this direction, as for example the Ophiuchus region shows significantly lower disk masses than the other SFRs even for (the albeit few) Class I systems \citep{Williams+2019,Cieza+2021}. 

The mass budget problem has thus always been discussed as potentially a timing issue, with early planet formation aligning well with the higher dust mass budget found in younger protoplanetary disks. \citet{SavvidouBitsch2025} show that on one hand early formation is crucial in order to form giant planets and on the other hand, the dust mass that gets trapped in the pressure bumps that giant planets generate is not accounted for following the flux-to-mass conversion and its corresponding assumptions and therefore, the estimated dust masses can be significantly underestimated but are then similar to the observed estimates. Even though these estimates end up in similar ranges to those from the surveys in various SFRs, planet formation has already happened in these models, thus showing that there is no mass budget problem for planet formation. 

Setting the mass budget problem aside, one of the main reasons why such emphasis is put on estimating accurate dust masses for the observed protoplanetary disks is that they serve as the missing piece to connect planet formation theories to the initial conditions in the natal environments. The amount of dust within a disk and how it evolves over time point to the availability of building blocks out of which planets can form and thus regulates how many and what types of planets can form in a given protoplanetary disk \citep[e.g.][]{Pollack+1996,KokuboIda1996,Mordasini+2009,Miguel+2011,Ronco+2017,Bitsch+2015,Bitsch+2019,Venturini+2020,Emsenhuber+2021,SavvidouBitsch2023}. It is thus crucial to verify whether the current assumptions for the initial conditions in planet formation theories are robust and at the same time whether the end products of the models are in good agreement with the data from the observed protoplanetary disks. 

The goal of the work here is two-fold: On one hand to show how the CDFs compare to the observations separated by the initial disk masses and to which degree the underestimation in the dust mass from the optically thin estimate comes from the dust that giant planets trap and on the other hand to show how the formation of larger bodies in general, e.g. planetesimals, will contribute to the missing mass problem given that they are currently observationally unattainable. The first part is discussed using the same sample as in \citet{SavvidouBitsch2023,SavvidouBitsch2025}, while for the second part, planetesimal formation is including in the model for a few example cases to show how the dust mass evolves in disks with and without efficient planetesimal formation.

\section{Methods}

The numerical simulations of planet formation in a protoplanetary disk are performed using the 1D semi-analytic \texttt{chemcomp} code \citep{SchneiderBitsch2021a} and include pebble growth and drift \citep{Birnstiel+2012}, pebble evaporation and condensation at ice lines \citep{SchneiderBitsch2021a}, planet growth via pebble \citep{JohansenLambrechts2017} and gas accretion \citep{Ndugu+2021}, and planet migration \citep{Paardekooper+2011}. The initial planetary mass of the embryos is determined by the pebble transition mass at which the planet starts efficient accretion from the Hill regime \citep{LambrechtsJohansen2012}. 

The model was described in detail in \citet{SavvidouBitsch2023} and \citep{SavvidouBitsch2025}. A parameter study was performed varying the disk mass ($M_{disk}$), disk radius ($R_{disk}$), $\alpha$-viscosity, the dust fragmentation velocity ($u_{frag}$), and, finally, the time ($t_0$) and location ($a_p$) of the inserted planetary embryo. Table \ref{tab:pars} shows the initial parameters used for the simulations (as discussed in \citep{SavvidouBitsch2023}). I note again here that all combinations of these parameters have been tested, for two dust-to-gas ratio values ($f_{DG}$=0.015 and $f_{DG}$=0.03), resulting in 
a total of 76800 runs. The stellar mass in all simulations is $M_\star~=~1~M_\odot$. The surface density of the disk exponentially decays at 3 Myr assuming that at this point, photoevaporation significantly depletes the disk from the gas \citep{Mamajek2009}.

\begin{table}[]
\centering
\begin{tabular}{@{}ccc@{}}
\toprule[1.2pt]
Parameter               & Values               &       \\                    
\midrule
$M_0$ {[}$M_{\odot}${]}  & 0.01, 0.04, 0.07, {\bf 0.1}   & initial disk mass            \\ \addlinespace
$R_0$ {[}$R_{\odot}${]}  & 50, 100, 150, {\bf 200}   & initial disk radius          \\ \addlinespace
$\alpha$                & {\bf 0.0001}, 0.0005, 0.001 & $\alpha$-viscosity parameter \\ \addlinespace
$t_0$ {[}Myr{]}          & {\bf 0.1}, 0.5, 0.9, 1.3   & starting time of embryo      \\ \addlinespace
$\alpha_{p,0}$ {[}AU{]} & 1-50 every 1  & initial position of embryo   \\ \addlinespace
$u_{frag}$ {[}m/s{]}    & 1, 4, 7, {\bf 10}    & fragmentation velocity     \\ \addlinespace
$f_{DG}$                & {\bf 0.015}, 0.03 & dust-to-gas ratio    \\ \bottomrule[1.2pt]
\end{tabular}
\caption{Parameters used in the simulations. We mark in bold the standard set, which is used as a reference in Fig. \ref{fig:Evolution_w_planetesimals}.}
\label{tab:pars}
\end{table}

The "optically thin" dust mass is calculated assuming that the optical depth $\tau$ is less than 1 in this case. Given also that 
\begin{equation} 
\tau = \Sigma_{dust} \cdot \kappa_{dust}~,
\end{equation}
the surface density that corresponds to the optically thin disk regions in the models is assumed to be
\begin{equation}
\Sigma_{dust,\tau<1} \leq 1/\kappa_{dust}.
\end{equation}
The common assumption of $\kappa_{dust}$ = 2.3 $cm^2/g$ at 1.3 mm is adopted \citep{Beckwith+1990}, as is done in flux-to-mass conversions from observations but it should be noted that changing the assumed opacity to any reasonable average value affects the results only minimally.

\begin{figure*}
    \centering
    \includegraphics[width=\textwidth]{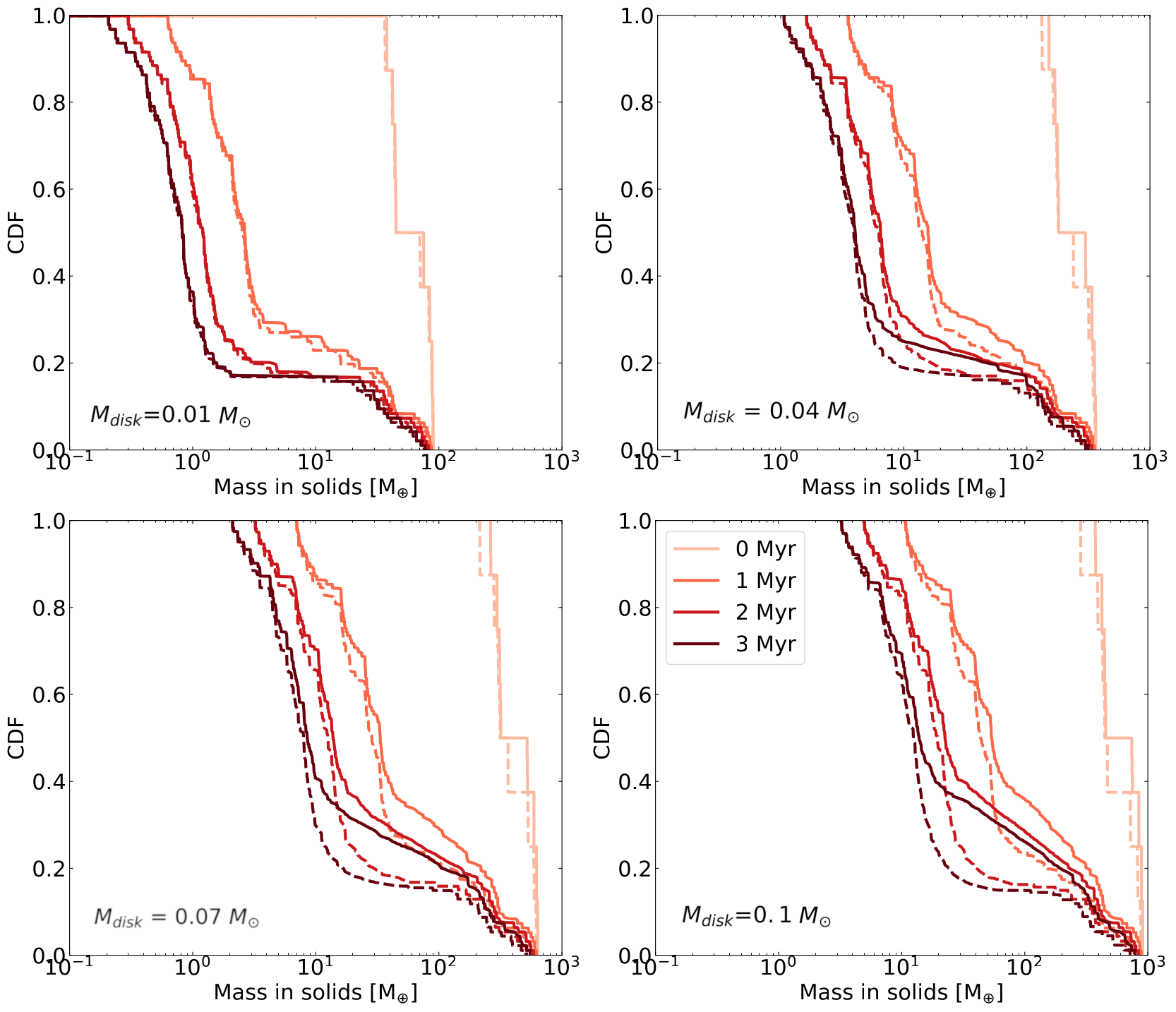}
    \caption{Cumulative distribution functions for the disk dust mass of our models at different times (0-3 Myr from lightest to darkest red), for each initial disk mass. All combinations of the rest of the parameters are included. The solid lines correspond to the total dust mass in our simulations, while the dashed lines correspond to the "optically thin" dust mass.}
    \label{fig:CDF_Mdust_M0}
\end{figure*}

\begin{figure*}
    \centering
    \includegraphics[width=\textwidth]{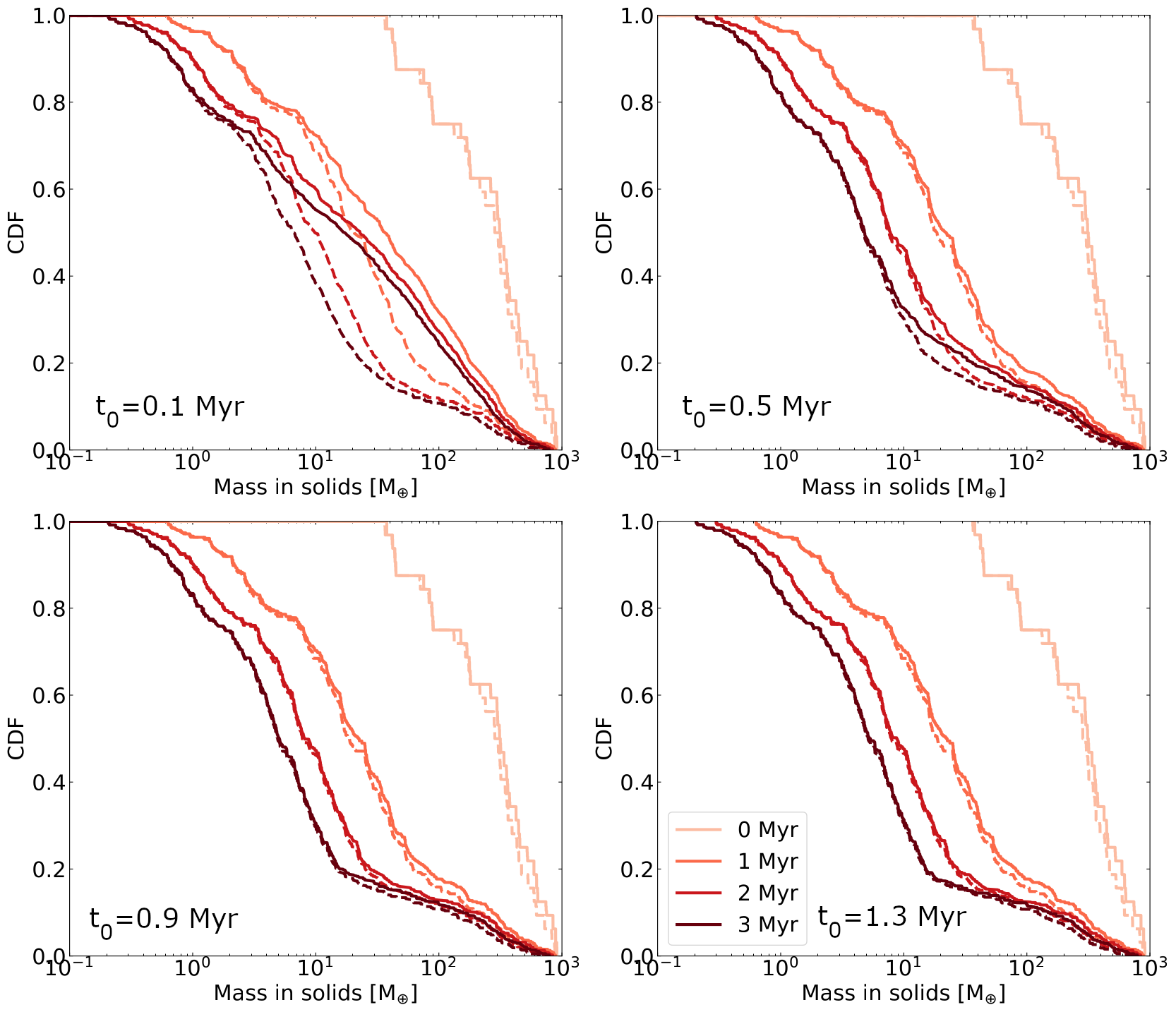}
    \caption{Cumulative distribution functions for the disk dust mass of our models at different times (0-3 Myr from lightest to darkest red), for each embryo injection time separately. All combinations of the rest of the parameters are included. The solid lines correspond to the total dust mass in our simulations, while the dashed lines correspond to the "optically thin" dust mass.}
    \label{fig:CDF_Mdust_t_0}
\end{figure*}

Planetesimal formation is not included in the full sample (Figs. \ref{fig:CDF_Mdust_M0} \& \ref{fig:CDF_Mdust_t_0}) which is the same as in \citet{SavvidouBitsch2023,SavvidouBitsch2025} but it is included in the example models discussed here (Fig. \ref{fig:Evolution_w_planetesimals}). It is modeled after the prescription in \citet{DrazkowskaAlibert2017}, with the condition for the Stokes number updated according to the results of \citet{LiYoudin2021}. Specifically, planetesimals form in the protoplanetary disk region where the Stokes number of pebbles exceed $10^{-3}$ and the dust-to-gas ratio exceeds 1. The rate of pebble to planetesimal conversion follows 
\begin{equation}
\dot{\Sigma_{pla}} = \zeta \cdot \Sigma_{dust} \cdot \Omega_K~,
\end{equation}
where $\zeta = 10^{-3}$ is the planetesimal formation efficiency. More details about this module of \texttt{chemcomp} can be found in \citet{Andama+2024}. In contrast to that work, the fragmentation velocity here is 10 m/s throughout the disk in all of the models shown. The nominal set of parameters is shown in bold in Table \ref{tab:pars}. I also show a case with higher $\alpha$, low disk mass and late embryo injection time, to discuss how the formation or not of a giant planet, along with potential planetesimal formation influences the dust mass time evolution. 

\section{Results}

\subsection{Dust mass evolution depending on the initial conditions}

In \citet{SavvidouBitsch2025}, we presented the inverse cumulative distribution functions (hereafter CDFs) for the whole sample of our models, from the beginning of the simulations until 3 Myrs, which is near the end of the lifetime of the disk in our models. Here, I present the CDFs of the same sample separated by the initial conditions that are the most critical for giant planet formation \citep{SavvidouBitsch2023}, specifically, the initial disk mass and the time of the embryo injection.

\subsubsection{CDFs per initial disk mass}

In Fig. \ref{fig:CDF_Mdust_M0}, I separate the CDFs by the initial disk masses, from $M_{disk}$ = 0.01 $M_{\odot}$ (top, left plot) to $M_{disk}$ = 0.1 $M_{\odot}$ (bottom, right plot). The solid lines correspond to the total dust mass in the models, while the dashed lines correspond to the estimated dust mass assuming optically thin emission and the flux-to-mass approximation of \citet{Hildebrand1983}. The difference between the total and the "optically thin" dust mass is negligible for the disks with the lowest mass (1\% $M_{\odot}$), while it increases with the increasing disk mass. We show in \citet{SavvidouBitsch2023} that giant planets predominantly form in high mass disks and there is thus a positive correlation between giant planet formation and disk mass. Therefore, as the disk mass increases, giant planets are more probable to form and also reach higher masses because of the faster and more efficient growth both of their cores via pebble accretion and also their envelopes via gas accretion. This leads to stronger dust traps and thus more dust mass that is unaccounted for in the "optically thin" approximation. 

The shape of the "optically thin" dust mass evolution is similar for all disk masses, but the overall dust mass increases with increasing disk mass. Keeping in mind the estimated cumulative distribution of mass around the observed young stars (e.g. Fig. 2 in \citet{Drazkowska+2023}), the highest dust mass in disks with $M_{disk} \leq 0.04~M_{\odot}$ (<300 $M_{\oplus}$) seems too low to match the maximum estimated ones for Class 0 observed disks. The best match for the Class 0 disks comes from the models that start with $M_{disk} = 0.07~M_{\odot}$, considering the dust mass at their earlier stages (0-1 Myr). Disks that start with very low disk mass (0.01 $M_{\odot}$) have at most around 90 $M_{\oplus}$ at the beginning of the disk evolution in the models. This would be in contrast to the observed protoplanetary disks that show maximum values of around 1000 $M_{\oplus}$ \citep[e.g.][]{Tychoniec+2020}. On the other hand, very massive disks maintain significantly high masses even after 3 Myr of evolution and planet formation (more than 10\% of the disks in this case have masses in solids above 100 $M_{\odot}$ at 3 Myr, even for the "optically thin" estimates). Therefore, intermediate initial disk masses seem to be more consistent with the observed population in regards to their initial and final masses. Even in this case, the fraction of disks starting with $M_{disk} = 0.07 M_{\odot}$ with dust mass at 3 Myr that is higher than 100 $M_{\oplus}$ is almost 20\%, while the maximum estimated dust mass in observed Class II disks does not reach this value. This discrepancy points to more lost mass by that point or more sources of underestimation, however, a direct comparison with the observations is not trivial. 

\subsubsection{CDFs per embryo injection time}

In Fig. \ref{fig:CDF_Mdust_t_0}, the CDFs are instead separated by the time of the embryo injection. The earliest injection time of $t_0$ = 0.1 Myr leads to the largest difference between the total dust mass in the models and the "optically thin" approximation (top, left plot). There is still a small difference with the embryo injection time of $t_0$ = 0.5 Myr (top, right plot) but as giant planet formation becomes limited for embryos that start growing later than this \citep{SavvidouBitsch2023}, the discrepancy in the CDFs is negligible (bottom plots). This once more clearly shows that the difference between the total dust mass and the "optically thin" approximation is caused by the optically thick dust mass that gets trapped in the giant planet induced pressure bumps. 

The cases of later embryo injection ($t_0$ = 0.9 and 1.3 Myr) lead to negligible differences and one could argue that this leaves the possibility open that the estimated dust masses in observed sources are robust and correspond to the total mass in solids in these disks. However, the fraction of giant planets that formed in these cases is almost zero with the nominal dust-to-gas ratio (1.5\%) and less than 10\% with an enhanced value (3\%), therefore it is contradicting the occurrence rates of giant planets over a wide range of orbital distances \citep[see discussion in][]{Drazkowska+2023}. This just points to the fact that some underestimation of the dust mass should be expected by the optically thick regions of dust in the disks. 

\subsection{Dust mass evolution when planetesimals form}

In the previous sections, I discussed how the dust masses in the models compare to those inferred from observations depending on the disk masses or the time when planets form. It was shown, already in \citet{SavvidouBitsch2025}, that the models are broadly consistent with the observations. Here it is further shown that taking into consideration the occurrence rates of giant planets as well as the constraints from the observed protoplanetary disk dust masses, the best matches come from intermediate mass disks and early planet formation times. The "optically thin" dust estimates are much lower than the total dust mass in these cases but they are still higher than those of Class II disks. 

Another potential reason why the estimated dust masses from observations seem to be too low for planet formation could be the dust and pebble mass that is converted to planetesimals. Such larger bodies cannot be observed, therefore the mass is just "lost" after planetesimal formation. While we did not include planetesimal formation in the original simulations of \citet{SavvidouBitsch2023} and \citet{SavvidouBitsch2025}, I show here some additional example cases in which planetesimal formation is included to discuss how the dust mass evolves with and without efficient planetesimal formation and how planet formation compares in these cases. 

\begin{figure*}
    \centering
    \includegraphics[width=\textwidth]{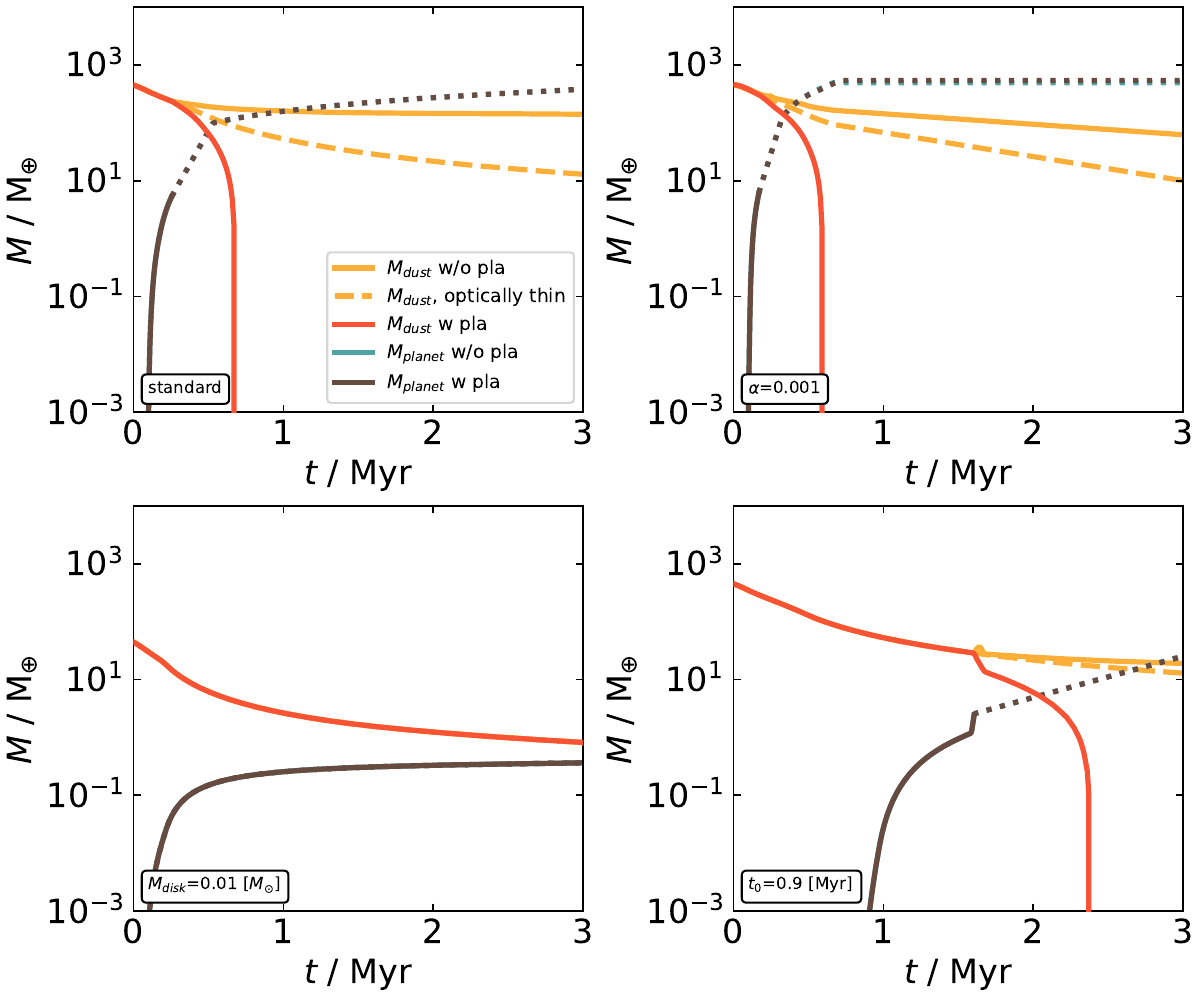}
    \caption{Mass (dust and planetary) as a function of time for a planet that started growing at 3 AU. The standard initial conditions are shown in bold in Table 
    \ref{tab:pars}. In each panel only one parameters changes. The yellow lines show the dust mass evolution in a model without planetesimal formation, with the solid lines corresponding to the total and the dashed lines corresponding to the optically thin estimate. The red lines show the dust mass evolution in a model with planetesimal formation. The gray and blue lines show the planetary mass time evolution for a model with and without planetesimal formation. Note that the gray line overlaps almost entirely with the blue line because the planetesimals mostly form after the planets reach the pebble isolation mass and thus their growth is not influenced.}
    \label{fig:Evolution_w_planetesimals}
\end{figure*}

Fig. \ref{fig:Evolution_w_planetesimals} shows how the dust mass evolves in a model with (red lines) and without planetesimal formation (yellow lines, solid for total mass and dashed lines for the optically thin estimate). I also overplot the time evolution of the planetary mass in each model (gray line for the model with planetesimal formation and blue for the nominal models without planetesimal formation). The planetary embryo started growing at 3 AU in all examples. \citep[see also][]{Danti+2023}. The top, left plot shows the standard set of parameters (bold in Table \ref{tab:pars}), while each other panel has only one parameter changed each time. The planetary mass is almost the same in all cases, thus regardless of whether planetesimal formation is included or not (therefore, the gray line in Fig. \ref{fig:Evolution_w_planetesimals} overlaps with the blue line). 

In all of the models shown in this example, planetesimals form mainly at the pressure bump that giant planets generate which is known to be a prime location for planetesimal formation \citep[e.g.][]{Eriksson+2020}. This leads to a rapid decrease of dust mass after the planet reaches the pebble isolation mass. These would be a second generation of planetesimals following already formed planets in the disk.
In the model with $\alpha$ = 0.001 (top, right plot) planetesimals initially form at the water iceline. In the model with the low disk mass (bottom, left plot), the planet does not reach the pebble isolation mass and, as a result, there is no difference in the dust mass evolution, with or without planetesimal formation. When the planetary embryo injection is delayed (bottom, right plot), then planetesimal formation is also delayed, until the planet reaches the pebble isolation mass. Consequently, the dust mass remains high for longer. 

More details on the conditions that enhance planetesimal formation can be found in \citet{Andama+2024}. It is shown here that while allowing planetesimals to form will not influence planet formation and a giant planet can still form efficiently, the dust mass will significantly decrease after planetesimals form. Therefore, the conversion to larger bodies which are then observationally unattainable can be another reason why especially Class II disks seem to have too low masses to support planet formation. 

\section{Discussion}

\subsection{The mass budget problem}
\label{Sect:mass_budget_discussion}

The reasons why there is no mass budget problem of planet formation were extensively discussed in \citet{SavvidouBitsch2025}. Briefly summarizing here, the reasons for the low derived masses in observed disks can be:
\begin{itemize}
\item Underestimation of disk masses from observations:
\begin{itemize}
\item Uncertainties in dust opacities and temperatures 
\item Optically thick emission, especially single wavelength dust continuum observations might be not best suited for accuracy
\item Non-interstellar medium like and non-constant dust-to-gas ratios 
\item Difficulty in obtaining robust gas masses 
\end{itemize}
\item Planet formation starts as early as possible (at least ongoing in Class II disks) 
\item Dust mass depletion (especially in the outer disk regions) over time due to radial drift and/or photoevaporation which truncates the disks after a few Myr
\item Dust traps (e.g. pressure bumps often created by giant planets or photoevaporation induced gaps) that are optically thick 
\item Boulders and planetesimals that form in disk are not observationally accessible and can thus contribute to the "missing" mass in evolved protoplanetary disks
\item Planet formation can be efficient even without the need for the highest disk masses (especially considering that not all stars have to make giant planets) 
\item The observed disks and exoplanets are biased and not necessarily the same populations, so comparison between them is not trivial
\end{itemize}

The majority of these points are discussed through the models presented and discussed in \citep{SavvidouBitsch2025} and here. 
Tying both of the points discussed in the previous section, \citep{Godines+2025} show that the dust clumping occurring at the early stages of planetesimal formation can be optically thick and lead to significant dust mass underestimations (factors of 2-7). They highlight, though, that the main factor of dust mass underestimation seems to be coming from the initial disk conditions and the impact of scattering on the (sub-)mm flux densities \citep{Zhu+2019,Liu2019,SierraLizano2020}. For example, considering porous dust opacities instead of the standard compact may alleviate the apparent lack of sufficient dust mass for planet formation \citep{Liu+2024}.

A comparison between the occurrence rates of structured protoplanetary disks in a large sample ALMA study and those of exoplanets (from radial velocity and transit surveys) in \citet{vanderMarelMulders2021} supports an evolutionary scenario where the initial disk mass determines whether a giant planet can form in the disk which is one of the mechanisms that can create pressure bumps that can potentially be observed in the millimeter. The results shown here but also in previous work \citep{SavvidouBitsch2023, SavvidouBitsch2025} are also consistent with this evolutionary scenario and show that these substructures in disks lead to significant underestimation of the dust mass but ultimately the low dust mass estimated from observations not only do not constitute a mass budget problem for planet formation but are also expected from planet formation theory and the pebble drift framework. 

\citet{Appelgren+2023} perform disk population synthesis models and show that the depletion of the dust mass due to radial drift can have a good agreement with the depletion trend seen in observations and a better agreement in the CDFs would come from including some mass retention in the disks with the highest masses. This is the case in our models, which include planet formation. The mass retained in disks with favorable conditions for giant planets ($M_{disk} \geq 0.04~M_\odot$) allowed for higher maximum dust mass (> 100 $M_\oplus$) in the CDFs that match well the observed ones. However, the fraction of disks with a dust mass higher than 100 $M_\oplus$ remains significantly higher compared to Class II sources \citep[e.g.][]{vanderMarelMulders2021} which show depletion at these stage. This discrepancy can decrease by considering optically thick regions, such as the ones created by trapped dust in planetary-induced pressure bumps, in which the dust mass is underestimated under the common methods. Further decrease of the dust mass could be caused by late-stage (post-planet formation) planetesimal formation at the same locations. Even though the efficiency remains uncertain, some amount of planetesimals is generally expected to form in pressure bumps \citep[e.g.][]{Johansen+2014,Eriksson+2020,Carrera+2021} and in this case, a significant amount of dust mass is converted to planetesimals, thus we should expect a reduction in the mm-emission. 

While the traditional method for estimating the dust masses of observed disks has been converting the (sub-)millimeter fluxes in the recent years, another method could be SED analysis, especially of multi-wavelength, continuum observations of the dust thermal emission \citep[e.g.][]{Xin+2023,Viscardi+2025}. However, obtaining multi-wavelength observations is limited by differences in the sensitivities and resolutions of the various facilities that operate at the different frequencies. Additionally, obtaining accurate dust mass measurements through SED analyses remains elusive, especially without the inclusion of optically thin tracers \citep{Viscardi+2025}. Nevertheless, further work in this direction could open up more robust pathways for accurate protoplanetary disk dust mass estimates.

\subsection{Planetesimal formation model caveats}

In this work, I expand on the discussion of \citet{SavvidouBitsch2025} about a few of the potential reasons behind the disk mass budget problem of planet formation that were summarized in the section above. In the first part that discusses the sample of \citet{SavvidouBitsch2023, SavvidouBitsch2025}, planetesimal formation is not included, therefore its influence in the CDFs is not discussed. The efficiency of planetesimal formation is still unknown and in order to discuss planet formation, planetesimal accretion would also be needed. 

The model used here follows the prescription by \citet{DrazkowskaAlibert2017} with the updated criterion by \citet{LiYoudin2021}, therefore planetesimals only form in specific locations that fulfill the criteria, in this work that is the water iceline and mainly the pressure bumps that giant planets create. However, other planetesimal formation models allow for formation in other locations, e.g. distributed at `pressure traps' over the whole disk \citep{Lenz+2019}, in which case a massive buildup of planetesimals at a specific location does not happen by default and a higher amount of dust can remain in the disk for longer. This could also differentiate between early-stage planetesimals, through which planetary embryos will emerge and planet formation will begin and late-stage planetesimals, forming after and potentially as a consequence of planets that have already grown and potentially reached the pebble isolation mass. It should be noted, that neither the final masses of giant planets nor the dust mass evolution are still significantly influenced when early-stage planetesimals are allowed to form \citep{Danti+2023}. This also means that the conclusion that intermediate initial disk masses seem to match the observations better could still be consistent with a population synthesis model that included planetesimal formation. 


Here, the pebbles exterior to the planetary gap (or water iceline initially) eventually get trapped there and in the case with planetesimal formation they get converted to planetesimals \citep{Eriksson+2020}. Additionally, pebbles get perfectly trapped there, however the extend to which dust gets filtered through the gap is an ongoing debate \citep[e.g.][]{Drazkowska+2019,Haugbolle+2019,Weber+2018}. Their fate after formation could also influence the dust mass evolution, as a significant part of their mass could be converted back to dust due to ablation, especially within the inner 10 AU \citep{Eriksson+2021}. Therefore, the dust depletion shown here could be an overestimation of how much dust mass can get converted to planetesimals. Nevertheless, the point to be made here is that dust conversion to planetesimals is one of the reasons for dust depletion, especially in evolved disks and planet formation could have already happened. 

\section{Conclusions}

The initial CDF of the models presented here starting with high or low disk masses (0.1 or 0.01 $M_\odot$) features too high ($\sim$ 900 $M_\oplus$) or too low ($\sim$ 90 $M_\oplus$) dust masses respectively in comparison to the masses of Class 0 disks \citep{Tychoniec+2020}. Taking this into consideration, the best match of the models with the estimated dust masses of observed protoplanetary disks comes from a disk mass of 0.04-0.07 $M_{\odot}$ that has produced a significant fraction of giants. In the whole sample models presented here, the low dust mass in the evolved disks is a direct consequence of (giant) planet formation and pebble radial drift. Interestingly, the fraction of giants that formed in disks with initial mass 0.07 $M_{\odot}$ is around 20\%, therefore in good agreement with the occurrence rates of giant planets \citep{Hsu+2020,Mulders+2018,Fulton+2021,Rosenthal+2021,Drazkowska+2023}. This also opens up the discussion of what an appropriate initial protoplanetary disk population would be and pointing to the fact that not all disks need to be initially heavy to succesfully form planets because not all disks have to form giant planets and/or multiple planets and even intermediate disk masses can form giant planets efficiently.

In general, the higher the initial disk mass, the larger the difference is between the total dust mass in the models and the "optically thin" dust mass. Similarly, the earlier embryos are injected into the disk models, the larger the discrepancy and in this case, it is almost negligible for injection times beyond 0.5 Myr. This is a direct consequence of giant planet formation in the models. Higher disk masses benefit giant planet formation which in turn generates a pressure bump and halts the inwardly drifting pebbles. The more mass that is trapped in a pressure bump means more mass that is not accounted for in the "optically thin" estimate. As a conclusion, giant planets and the optically thick regions they create can be one of the main solutions for the mass budget problem of planet formation \citep{SavvidouBitsch2025}. However, even the "optically thin" dust masses at 3 Myr are higher compared to Class II disks \citep[e.g.][]{vanderMarelMulders2021}.

Under fruitful conditions a significant amount of dust mass can get converted to planetesimals \citep[e.g.][]{Johansen+2014,Eriksson+2020,Voelkel+2022,Andama+2024} and this will lead to another decrease in the observed dust mass that can alleviate this difference. It is also shown here that especially after planet formation, the pressure bumps that planets generate are prime locations for planetesimal formation and, in this case, the dust mass decreases then by several orders of magnitude rapidly. Giant planet formation is not affected, as these planetesimals in this work form after the planet reaches the pebble isolation mass but even when they are allowed to form from the beginning, the final planetary masses and the dust evolution remain similar 
\citep{Danti+2023}. Therefore, even though the apparent dust mass would seem very low if such a disk was observed, planets and planetesimals could be already formed, and thus even very low disk masses would not pose a problem to planet formation. The observed sources, and thus even disks that have already formed giant planets, correspond to a population spanning a wide range of phases in their evolution. This means that on a population level, some of the observed disks could contain an amount of trapped dust (which could be also caused by other mechanisms) and some in later stages might contain a significant amount of (a post-planet formation generation of) planetesimals. 

The models presented in \citep{SavvidouBitsch2025} and here show that the current theoretical framework and including self-consistently most of the factors discussed in Sect. \ref{Sect:mass_budget_discussion} leads to consistent dust mass evolution to observations. While beyond the scope of this paper, future work will address planetesimal formation as well on a population level to allow for appropriate comparisons with the observed protoplanetary disk sample and their estimated masses.

More work needs to be done in understanding the earliest stages of planet formation alongside the formation of the young stars they orbit and the conditions that constitute the natal environments of the forming planets. Additionally, alternative methods, even if more complex or expensive, should be considered for obtaining robust dust mass or even disk/gas mass estimates from observations. However, even at this stage, the disk mass budget problem seems to be highlighting our shortcoming in reconciling our theoretical understanding with the observational data more than it is an issue with the standard framework of planet formation. 

\section*{Acknowledgments}
I would like to thank the two referees who helped to improve the quality of the manuscript. S.S. acknowledges support from the National Aeronautics and Space Administration (NASA) awards for JWST-GO-01584 and JWST-GO-01549.

\bibliographystyle{Frontiers-Harvard} 
\bibliography{PaperV}

\section*{Figures}

\end{document}